\documentclass[aps,pra,floatfix,showpacs,superscriptaddress]{revtex4-1}
\usepackage[colorlinks=true,linkcolor=blue]{hyperref}
\usepackage[utf8]{inputenc}
\usepackage{siunitx,calc,graphicx,amsmath,amssymb,bm,natbib,booktabs,float,caption,subcaption,enumitem,ragged2e,tabularx,multirow,xcolor,adjustbox,array}
\newcommand{\Omegam}{\Omega_{\mathrm{m0}}}

\newcommand{\Omegar}{\Omega_{\mathrm{r0}}}
\newcommand{\omegam}{\omega_{\mathrm{m}}}
\newcommand{\omegab}{\omega_{\mathrm{b}}}
\newcommand{\omegar}{\omega_{\mathrm{r}}}

\newcommand{\oo}{\omega_0}
\newcommand{\wa}{\omega_a}
\newcommand{\wpp}{\omega_p}
\newcommand{\LCDM}{\Lambda\mathrm{CDM}}
\newcommand{\wCDM}{\omega\mathrm{CDM}}
\newcommand{\wowaCDM}{\omega_{0}\omega_{a}\mathrm{CDM}}

\begin{document}
\title{Assessing the Robustness of the CPL Parametrization to Basis and Prior Variations: Insights from DESI DR2 BAO Data}
\author{Seokcheon Lee}
\email{skylee@skku.edu}
\affiliation{Department of Physics, Institute of Basic Science, Sungkyunkwan University, Suwon 16419, Korea}
\date{\today}

\begin{abstract}
This work reexamines cosmological parameter constraints from the DESI Data Release~2 baryon acoustic oscillation (BAO) measurements using the distance-basis representation $(D_V/r_d,\,D_M/D_H)$, which separates the isotropic BAO scale from the scale-free Alcock–Paczynski ratio. This work compares $\Lambda$CDM, $\wCDM$, and $\wowaCDM$ models to evaluate how the choice of data basis and the width of the prior on $\wa$ affect dark-energy inference. 
Ratio-only fits ($D_M/D_H$) amplify the $(\oo,\,\wa)$ degeneracy and can produce large apparent shifts in point estimates without genuine evidence for dynamical dark energy. 
Joint fits using $(D_V/r_d, D_M/D_H)$ restore parameter consistency and show that these shifts mainly trace the degeneracy ridge. 
The pivoted equation of state, $\omega_p=\omega(a_p)\simeq-0.9\pm0.1$ at $z_p\simeq0.34$, remains stable and consistent with a cosmological constant within $1\sigma$. 
Model-selection diagnostics (AIC, BIC, and Bayes factors) provide only moderate support for $\Lambda$CDM, indicating no significant evidence for an evolving $\omega(a)$. 
These findings clarify the interplay among basis choice, absolute-scale anchoring, and degeneracy geometry in BAO-only dark-energy analyses, providing a benchmark for future DESI and next-generation surveys.
\end{abstract}

\maketitle

\section{Introduction}
\label{sec:intro}

Baryon acoustic oscillations (BAO) provide one of the most precise geometric probes of cosmic expansion, serving as a standard ruler that links late-time distance measurements to early-universe physics~\cite{SDSS:2005xqv,BOSS:2013rlg,BOSS:2016wmc,eBOSS:2020lta}.  
Modern galaxy surveys such as the Dark Energy Spectroscopic Instrument (DESI) have now achieved sub-percent precision in both transverse and radial BAO measurements, enabling stringent tests of the $\LCDM$ model and its extensions~\cite{DESI:2024mwx,DESI:2025fxa,DESI:2025zgx,DESI:2025fii}.  
Because BAO observables are derived from the correlation of comoving distances and the Hubble scale, their interpretation depends on how the underlying distance ratios are expressed and on whether the absolute sound-horizon scale $r_d$ is treated as an external prior or a self-consistent variable.

Recent joint analyses that combine the DESI~DR2 BAO data with external probes such as the cosmic microwave background (CMB) and type~Ia supernovae (SNe Ia) have reported apparent deviations from a cosmological constant, often interpreted as hints of dynamical dark energy (DDE)~\cite{Wang:2024qan,DESI:2025wyn,Cortes:2025joz,Liu:2025mub,Ishak:2025cay,Chaudhary:2025vzy,Chaudhary:2025bfs}.  In these combined fits, the inclusion of strong external priors on $\Omegam$, $h$, or $r_d$ compresses the posterior along the $(\oo,\,\wa)$ degeneracy ridge, typically driving $\wa$ toward negative values and producing the illusion of an evolving equation of state.  While such results have been cited as evidence for DDE, their statistical significance remains limited and may largely reflect the geometry of parameter degeneracies rather than new physical dynamics.  This motivates a reassessment of what the BAO data alone truly imply about DE evolution, independent of external anchors or prior-induced correlations~\cite{Colgain:2024mtg,Sakr:2025daj,Jiang:2025ylr,Huang:2025som,Lee:2025kbn,Chaudhary:2025pcc,Toomey:2025xyo}.

Most previous BAO-only analyses have adopted the pair of observables $(D_M/r_d,\, D_H/r_d)$, or equivalently $(\alpha_\perp,\alpha_\parallel)$, which separately trace the transverse and radial dilation relative to a fiducial model~\cite{SDSS:2005xqv,Kazin:2014qga}.  While convenient, this representation can lead to information redundancy and potential double-counting when combined with scale-free ratios such as $D_M/D_H$ or $D_V/r_d$, because these quantities are algebraically related and share overlapping covariance information.  To avoid this issue, the DESI collaboration reports their official constraints in the $(D_V/r_d,\, D_M/D_H)$ basis, where $D_V$ encapsulates the isotropic BAO scale and $D_M/D_H$ encodes the anisotropic Alcock–Paczynski  (AP) ratio.  This distance-basis representation therefore provides a non-redundant and internally consistent description of the BAO signal, separating its absolute and relative geometric contributions.

Beyond the choice of data basis, model comparison among $\LCDM$, $\wCDM$, and $\wowaCDM$ is strongly influenced by parameter degeneracies.  
In the Chevallier–Polarski–Linder (CPL) parametrization~\cite{Chevallier:2000qy,Linder:2002et}
\begin{equation}
\omega(a) = \oo + \wa (1 - a), \label{CPL}
\end{equation}
the DE equation-of-state (EoS) parameters $\oo$ and $\wa$ are highly anti-correlated because most low-redshift observables constrain a nearly constant combination of them (\textit{i.e.}, $\oo+\wa \simeq$ const.).  
This yields a narrow degeneracy ridge in the $(\oo,\,\wa)$ plane~\cite{Tegmark:1996bz,Albrecht:2006um,Huterer:2000mj,Linder:2006xb,Yang:2018prh}.  
Apparent shifts of $(\oo,\,\wa)$ along this ridge can occur with minimal change in likelihood, creating the misleading impression of time-varying DE even when the data remain fully consistent with $\omega=-1$~\cite{Huterer:2000mj,Albrecht:2006um,Linder:2006sv,Pan-STARRS1:2017jku}.  

The $\oo$–$\wa$ degeneracy has been extensively analyzed in the literature~\cite{Szydlowski:2006pz,Gong:2013bn,Shlivko:2024llw},  showing that most late-time probes constrain only a single effective combination of these parameters along an elongated ridge. As a result, broadening the prior on $\wa$ or adopting an alternative data basis primarily moves the posterior along this degeneracy direction rather than revealing genuine time evolution in $\omega(a)$.  Understanding this geometric correlation is crucial, since it can lead to apparent evidence for DDE that merely reflects parameter covariance rather than new physics.

Several studies have pointed out that the CPL parametrization, while convenient for phenomenological modeling, can exhibit artificial evolution or unphysical behavior at high redshift, thereby obscuring the true nature of DE and biasing cosmological inference~\cite{Weller:2001gf,Sahni:2006pa,Linden:2008mf}.  Furthermore, the CPL form’s linear dependence on scale factor imposes a restricted functional space that cannot capture non-monotonic or rapid variations in $\omega(z)$, 
and its pivot redshift is often dataset-dependent~\cite{Caldwell:2005tm,Kitching:2008vx,Scherrer:2015tra,Nesseris:2025lke,Ozulker:2025ehg}. These limitations motivate the present reanalysis, which examines whether apparent evidence for evolving $\omega(a)$ arises from data sensitivity or from the parametrization and prior geometry itself.

To isolate the physically constrained combination, one defines the pivoted EoS
\begin{equation}
\wpp = \omega(a_p) = \oo + (1-a_p) \wa , \label{wpp}
\end{equation}
where $a_p$ (or $z_p$) is the pivot scale factor at which the covariance $\mathrm{Cov}(\wpp,\,\wa)=0$~\cite{Astier:2000as,Huterer:2000mj,Linder:2006xb,Scovacricchi:2012fre}.  
This pivot formalism reprojects the likelihood into the direction of maximal information and minimizes prior dependence, yielding a more robust summary parameter than either $\oo$ or $\wa$ individually.

In practice, degeneracies between $(\Omegam,\,h r_d,\,\oo,\,\wa)$ also arise because BAO measurements alone are sensitive primarily to the dimensionless ratios $D_M/r_d$ and $D_H/r_d$.  
Unless the sound horizon $r_d$ is anchored by external data such as the CMB, the BAO likelihood becomes nearly scale-free, leading to elongated likelihood contours that follow the same Fisher-matrix eigenmode as the CPL ridge.  Hence, the physical interpretation of DE constraints critically depends on whether $r_d$ is fixed, marginalized, or jointly inferred from $(\omegab,\omegam,h)$~\cite{Lee:2025rmg}. In this work, to ensure a pure BAO-only constraint without external $r_d$ calibration, the combined parameter $h r_d$ is sampled directly.

The present work revisits DESI DR2 BAO data~\cite{DESI:2025zgx} to evaluate how cosmological parameter constraints depend on the adopted BAO basis and on the prior range of the evolution parameter $\wa$.  Three complementary configurations are considered. First, a mixed basis $(D_V/r_d,\,D_M/D_H)$ is adopted as our fiducial case, separating the isotropic absolute-scale information from the scale-free geometric ratio.  Second, a ratio-only analysis using $D_M/D_H$ isolates the purely geometric AP information, providing a diagnostic of scale-independent constraints.  Finally, the conventional DESI basis $(D_M/r_d,\,D_H/r_d)$ is examined for comparison with earlier analyses~\cite{Lee:2025rmg}, enabling direct assessment of how basis choice influences the inferred posteriors.  Each case is analyzed under both narrow and wide priors on $\wa$ to quantify the sensitivity of posterior inferences to prior volume and degeneracy geometry.

Markov Chain Monte Carlo (MCMC) sampling is performed with flat priors in $(\Omegam,\,h r_d,\,\oo,\,\wa)$, computing $\chi^2_{\min}$, AIC, BIC, and Bayes factors via the Savage–Dickey density ratio (SDDR)~\cite{Dickey:1971,Verdinelli:1995,Trotta:2005ar,Heavens:2007ka,Trotta:2008qt,Bridges:2008ta,Higson:2018cqj,Planck:2018vyg,Rezaei:2021qpq,DiValentino:2021izs,Wang:2025bkk,Lin:2025xbw}.  
Posterior covariances and pivot parameters are derived directly from the MCMC chains to quantify the degeneracy structure and to assess the robustness of inferred trends.
All model fits are performed self-consistently to ensure compatibility with the official DESI DR2 likelihood.

Recent studies have directly tested the internal consistency of $\LCDM$ itself~\cite{Liu:2024fjy}.  
For example,  the comparison between the early- and late-universe constraints on the physical matter density $\omegam$ using BAO and CMB data, finding ${\rm ratio}(\omegam)=1.0069\pm0.0070$, consistent with unity.  
Their result confirms that the apparent early–late discrepancies can arise without requiring any new physics, underscoring the importance of degeneracy-aware analyses like ours.

The remainder of this paper is organized as follows.  
To isolate the geometric compensation effects between the DE parameters and the matter density, the physical impact of the increased matter density that arises in CPL posteriors is first discussed in Section~\ref{sec:omegam_effect}.
Section~\ref{sec:data_model} summarizes the DESI DR2 BAO observables and model setup.  
Section~\ref{sec:methodology} outlines our statistical framework, including likelihood construction, parameter sampling, and model-comparison metrics.  
Section~\ref{sec:results} presents the posterior distributions, degeneracy analyses, and model-selection outcomes for Scenarios~A–C.  
Finally, Section~\ref{sec:conclusion} summarizes the implications of our findings, emphasizing that apparent variations in $(\oo,\wa)$ along the degeneracy ridge do not constitute evidence for DDE, and that the pivoted equation of state $w_p\simeq -0.9\pm0.1$ remains consistent with $\Lambda$CDM within current BAO precision.

\section{Effect of the Increased Matter Density in CPL Posteriors}
\label{sec:omegam_effect}

Before discussing the data and methodology in detail, it is instructive to examine a characteristic feature of the CPL ($\wowaCDM$) model's posterior distributions.  
In both DESI DR2 and combined BAO--CMB analyses, the CPL fits tend to favor noticeably higher present-day matter densities, $\Omegam\!\gtrsim\!0.34$, compared to the baseline $\LCDM$ value of $\Omegam\simeq0.30$.  
Such an increase may appear moderate, but it has nontrivial physical implications for both early- and late-universe observables.  
In particular, the associated rise in $\Omegam$ (or equivalently $\omegam = \Omegam h^2$) modifies the matter--radiation equality epoch and the comoving horizon scale, and when combined with a lower $H_0$ can exacerbate the Hubble-tension problem.  
Here, the ratio is defined as
\begin{equation}
\mathrm{ratio}(\omegam) \equiv 
\frac{\omegam^{\rm late}}{\omegam^{\rm early}},
\end{equation}
where $\omegam^{\rm late}$ is inferred indirectly from late-time distance data
under a given cosmological model, and $\omegam^{\rm early}$ is obtained from CMB
constraints. We emphasize that BAO data alone do not directly measure $\omegam$.
In a BAO-only analysis, the likelihood is sensitive primarily to the combinations
$H_0 r_d$ and dimensionless distance ratios.
Any discussion of $\omegam$ therefore serves as a derived, model-dependent
interpretation rather than a direct observable constraint. This section provides a quantitative illustration of these effects.

A notable feature of the $\wowaCDM$ fits to BAO data is their tendency to favor a higher present-day matter density $\Omegam$ compared to $\LCDM$ or $\wCDM$.  
This behavior is closely tied to the anti-correlation between $(\oo,\,\wa)$ along the CPL degeneracy ridge:  
as $\wa$ becomes more negative, the expansion rate at intermediate redshift is reduced, which can be compensated by increasing $\Omegam$ without significantly altering the overall $\chi^2$.  This inverse response between $\Omegam$ and $\wa$ therefore represents a purely geometric compensation effect rather than a new physical mechanism.  
Understanding this effect is important because changes in $\Omegam$ also propagate into early-universe quantities such as the redshift of matter–radiation equality and the comoving sound horizon at the drag epoch.

This geometric compensation effect is not limited to late-time dynamics. The resultant change in $\Omegam$ also propagates to early-universe quantities. The redshift of matter--radiation equality and the comoving horizon scale at that epoch are determined primarily by the physical matter density $\omegam$, rather than by $\Omegam$ or $h$ separately
\begin{align}
1+z_{\rm eq} &= \frac{\omegam}{\omegar}\,, \qquad
\omegar = \omega_{\gamma}\!\left[1+ \frac{7}{8} \left( \frac{4}{11} N_{\rm eff} \right)^{4/3}\right], \label{zeq}
\end{align}
where $\omega_{\gamma}\equiv\Omega_{\gamma0}h^2$.  
For $T_{\rm CMB}=2.7255\,{\rm K}$ and $N_{\rm eff}=3.046$, one obtains $\omegar\simeq4.177\times10^{-5}$, so that
\begin{equation}
z_{\rm eq}\simeq2.394\times10^{4}\,(\omegam)-1\, . \label{zeq2}
\end{equation}
If $\Omegam$ increases from $0.30$ to $0.35$ at fixed $h$, then $z_{\rm eq}\propto\Omegam$ and rises by roughly $+17\%$:  
for $h=0.67$, $z_{\rm eq}\approx3223\!\rightarrow\!3761$, and for $h=0.70$, $z_{\rm eq}\approx3518\!\rightarrow\!4105$.  
The corresponding comoving wavenumber at equality,
\begin{equation}
k_{\rm eq}\equiv\frac{a_{\rm eq}H_{\rm eq}}{c}\simeq\sqrt{2}\,\frac{H_0}{c}\,\frac{\Omegam}{\sqrt{\Omegar}}\approx0.0729\,\omegam\ {\rm Mpc}^{-1}, \label{keq}
\end{equation}
also increases by the same fraction.  For $h=0.67$, $k_{\rm eq}\!\approx\!9.8\times10^{-3}\!\rightarrow\!1.15\times10^{-2}\,{\rm Mpc}^{-1}$, corresponding to a turnover scale shift $\lambda_{\rm eq}=2\pi/k$ from $\sim640$ to $\sim549$\,Mpc ($-14\%$).  
Thus, even a modest change in $\Omegam$ significantly alters the early-time horizon scale. The numerical examples in Eq.~\eqref{keq} are chosen to illustrate typical parameter combinations encountered along the CPL degeneracy ridge in DESI DR2 BAO fits. They are not best-fit values, but representative points showing how variations in $(\Omegam, h)$ can leave the physical matter density $\omegam$ nearly unchanged.

In realistic joint fits with early-universe data, however, $\omegam$ is tightly constrained.  The small, yet non-zero, $\omegam$ shift that typically occurs when moving from $\LCDM$-like parameters to CPL-preferred high-$\Omegam$ solutions is illustrated. For example,
\begin{align}
&\Omegam=0.300,\ h=0.670 \Rightarrow \omegam=0.135,  \nonumber \\ 
&\Omegam=0.315,\ h=0.674 \Rightarrow \omegam=0.143,  \label{Omegam} \\
&\Omegam=0.353,\ h=0.636 \Rightarrow \omegam=0.143, \nonumber
\end{align}
corresponding to only a $+7.1\%$ increase in $\omegam$.  
In this case,
\begin{align}
z_{\rm eq}:&\ 3223 \rightarrow 3451 \quad (+7.1\%)\,, \qquad
k_{\rm eq}: 9.8\times10^{-3} \rightarrow 1.05\times10^{-2}\ {\rm Mpc}^{-1}\quad (+7.1\%)\label{zeqkeq} \,.
\end{align}
The sound horizon at the drag epoch scales as $r_d\propto\omegam^{-0.23}$~\cite{Brieden:2022heh}, hence a $+7.1\%$ rise in $\omegam$ reduces $r_d$ by only $\sim1.6\%$.  
CMB acoustic-peak positions and the turnover in the matter power spectrum remain nearly unchanged because they depend primarily on $\omegam$.  It is noted that the turnover-scale shift discussed here implicitly reflects the
change in the dimensionless Hubble parameter $h$ between $\Lambda$CDM and CPL
solutions.
At fixed $\omegam$, an increase in $\Omegam$ necessarily implies a lower $h$,
which contributes to the apparent shift in $k_{\rm eq}$ when expressed in
$\mathrm{Mpc}^{-1}$ units.  For reference, the turnover scale has also been directly measured by
the DESI collaboration and reported in units of $h\,\mathrm{Mpc}^{-1}$~\cite{Bahr-Kalus:2025hhb}.

At fixed $\omegam$, however, an increase in $\Omegam$ must be offset by a lower $h$ according to
\begin{equation}
h=\sqrt{\frac{\omegam}{\Omegam}}, \qquad H_0=100h \label{h} \,.
\end{equation}
Numerically, for $\omegam=0.143$,  
$\Omegam=0.300\Rightarrow H_0\simeq69.0$,  
whereas $\Omegam=0.352\Rightarrow H_0\simeq63.7$.  
Thus $H_0$ decreases by $\sim7.7\%$, moving farther below both the Planck value ($67.4$) and the SH0ES result ($\sim73$), thereby worsening the Hubble tension.

The physical matter density, $\omegam$, also determines the sound horizon via
\begin{equation}
r_d = \int_{z_d}^{\infty} \frac{c_s(z)}{H(z)}\,dz \,. \label{rd}
\end{equation}
A higher $\Omegam$ or $\omegam$ leads to an earlier equality epoch ($z_{\rm eq}$ larger) and a smaller sound horizon $r_d$.  
Consequently, $\wowaCDM$ models with elevated $\Omegam$ can mimic the same BAO observables as $\LCDM$ by compensating a reduced $r_d$ against changes in $H(z)$ at low redshift.  This degeneracy is purely geometric and intrinsic to distance-ratio measurements, meaning that the apparent shift in $\Omegam$ does not by itself signal any deviation from $\LCDM$ dynamics.

To assess the magnitude of this effect, $r_d$ is explicitly computed for each model using the DESI DR2 fitting formula~\cite{Brieden:2022heh}
\begin{equation}
r_d^{\rm DESI} = 147.05\,{\rm Mpc}
\left(\frac{\omegab}{0.02236}\right)^{-0.13}
\left(\frac{\omegam}{0.1432}\right)^{-0.23}
\left(\frac{N_{\rm eff}}{3.04}\right)^{-0.1}.
\label{eq:rd_fitting}
\end{equation}
This empirical relation, calibrated from Boltzmann-code computations, ensures that variations in $\Omegam$ or $h$ consistently modify $r_d$ through their impact on $\omegab$ and $\omegam$.
In practice, the apparent high--$\Omegam$ solutions primarily reflect the geometric degeneracy between $\omega(a)$ and $r_d$, rather than a genuine preference for higher matter content.  
Therefore, shifts toward larger $\Omegam$ in CPL posteriors should be regarded as movements along this compensating direction in parameter space, not as indications of an increased matter budget.

This phenomenon underscores the importance of performing joint analyses with complementary probes that anchor either $r_d$ or $H_0$.  However, as demonstrated in recent studies~\cite{Sakr:2025daj,Jiang:2025ylr,Huang:2025som,Lee:2025kbn,Toomey:2025xyo}, the inclusion of strong external priors—particularly those on $\Omegam$ from supernova or CMB data—can inadvertently shift the inferred $(\oo, \,\wa)$ values away from their true $\LCDM$ point.  
When the prior mean is even slightly inconsistent with the BAO-only constraint, the resulting posterior can mimic apparent evidence for DDE despite originating from a purely $\LCDM$ universe.  
Such prior-induced biases tighten the confidence contours while displacing their centers, giving a false impression of physical evolution.  
Therefore, multi-probe analyses must ensure prior consistency across datasets before interpreting deviations in $(\oo,\,\wa)$ as signatures of new physics.  Within a BAO-only context, the elevated $\Omegam$ values seen in CPL posteriors should thus be regarded as geometric artifacts of the near scale-free likelihood, not as evidence for genuine DE dynamics. This interpretation is consistent with the independent $\LCDM$ consistency test of Ref.~\cite{Liu:2024fjy},  
which demonstrated that early–late parameter variations in $\omegam$ remain fully compatible within 1-$\sigma$.  
Our finding that the CPL-induced $\Omegam$ shift traces a geometric degeneracy rather than new dynamics thus echoes their conclusion that parameter-level deviations do not necessarily signal physics beyond $\LCDM$.

\section{DESI DR2 BAO Data and the $\wowaCDM$ Model}
\label{sec:data_model}

The DESI~DR2 BAO observables are analyzed in the distance basis $(D_V/r_d,\,D_M/D_H)$, which separates the isotropic BAO scale from the scale-free Alcock--Paczynski (AP) ratio~\cite{DESI:2025zgx}.  
The isotropic combination is defined as
\begin{align}
D_V(z) &= \left[D_M^2(z)\,\frac{cz}{H(z)}\right]^{1/3}, \label{DV}
\end{align}
and the AP ratio as
\begin{align}
\frac{D_M(z)}{D_H(z)} &= E(z)\!\int_0^z\!\frac{dz'}{E(z')}, 
\qquad E(z)=\frac{H(z)}{H_0}. \label{DMoDH}
\end{align}
Although $(D_V/r_d)$ and $(D_M/D_H)$ are only weakly correlated in the DESI~DR2 measurements 
(the correlation coefficient is typically $|\rho|\!\lesssim\!0.1$ across all redshift bins), 
they are not strictly independent.  
In this work, the full covariance matrix provided in~\cite{DESI:2025zgx,Lee:2025kbn}
is therefore employed when constructing the Gaussian likelihoods used in the MCMC sampling.  
This treatment preserves the small residual correlations between $D_V/r_d$ and $D_M/D_H$ 
and ensures a statistically consistent inference across all BAO redshift bins.

\subsection{Background model}
For DE, the CPL parametrization is adopted,
\begin{align}
E(z) &= \Big[\Omega_r (1+z)^4 + \Omega_m (1+z)^3 
+ \Omega_{\rm de} (1+z)^{3(1+\oo+\wa)} 
e^{-3\wa z/(1+z)}\Big]^{1/2}, \label{Ez}
\end{align}
with spatial flatness $\Omega_{\rm de}=1-\Omega_m-\Omega_r$.  
Using the physical densities $\omega_i\equiv\Omega_i h^2$, the expansion rate may equivalently be written as
\begin{align}
E_h(z) = 
\Big[\omega_r (1+z)^4 + \omega_m (1+z)^3 
+ \omega_{\rm de} (1+z)^{3(1+\oo+\wa)}
e^{-3\wa z/(1+z)}\Big]^{1/2}, \label{Ehz}
\end{align}
where $\omega_{\rm de}=h^2-\omega_m-\omega_r$.  
Early-time parameters $(\omega_\gamma,\,N_{\rm eff})$ are fixed to Planck values. 

Because the BAO likelihood involves both the absolute-scale observable $D_V/r_d$ and the scale-free AP ratio $D_M/D_H$,  
it is necessary to treat the sound horizon consistently.  To perform a pure BAO-only analysis without relying on external CMB-derived priors on $r_d$,  the official DESI~DR2 BAO-only approach is followed~\cite{DESI:2025zgx}  and sample the combined quantity $h\,r_d$ independently in the MCMC, alongside $\Omegam$, $\oo$, and $\wa$.  This treatment constrains $r_d$ through the observable combination that BAO actually measures,  avoiding the imposition of any external absolute-scale prior while maintaining full consistency with the DESI likelihood construction.

\subsection{Mapping between BAO fitting parameters and distances}
The anisotropic BAO measurement is first obtained in terms of dilation parameters,
\begin{align}
\alpha_\parallel(z) &\equiv 
\frac{D_H(z)\,r_d^{\rm fid}}
     {D_H^{\rm fid}(z)\,r_d}, \qquad
\alpha_\perp(z) \equiv 
\frac{D_M(z)\,r_d^{\rm fid}}
     {D_M^{\rm fid}(z)\,r_d}.
\end{align}
From these, DESI defines
\begin{align}
\alpha_{\rm iso}(z)
  &= [\alpha_\parallel(z)\,\alpha_\perp^2(z)]^{1/3}, \qquad
\alpha_{\rm AP}(z)
  = \frac{\alpha_\parallel(z)}{\alpha_\perp(z)},
\end{align}
which map one-to-one onto the observables $(D_V/r_d,\,D_M/D_H)$:
\[
D_V/r_d \leftrightarrow \alpha_{\rm iso}, \qquad
D_M/D_H \leftrightarrow \alpha_{\rm AP}.
\]
Hence $D_V/r_d$ captures the isotropic, $r_d$-dependent BAO ruler, 
while $D_M/D_H$ isolates the anisotropic geometry through the AP ratio.  
Using this mixed basis ensures that the two observables jointly encode the complete BAO information without double counting or covariance overlap.

\begin{table}[t]
\centering
\caption{Mapping between BAO fitting parameters and distance observables.}
\begin{ruledtabular}
\begin{tabular}{lcc}
\toprule
Quantity & From $\alpha$-basis & From distances \\
\hline
\addlinespace[0.3em]
$D_V/r_d$ & $\alpha_{\rm iso}\,[D_M^{\rm fid\,2}zD_H^{\rm fid}/r_d^{\rm fid\,3}]^{1/3}$ & $[D_M^2zD_H]^{1/3}/r_d$ \\
$D_M/D_H$ & $(1/\alpha_{\rm AP})(D_M^{\rm fid}/D_H^{\rm fid})$ & $D_M/D_H$ \\
\bottomrule
\end{tabular}
\end{ruledtabular}
\label{tab:alpha_mapping}
\end{table}

\noindent
In summary, the $(D_V/r_d,\,D_M/D_H)$ basis cleanly separates the absolute-scale information from the relative geometry, aligning with the DESI DR2 analysis protocol and preventing redundancy among correlated distance measures.  
It further provides an ideal framework for diagnosing prior-dependent biases that can be misinterpreted as evidence for DDE.

\section{Methodology}
\label{sec:methodology}

Gaussian likelihoods are constructed from the DESI DR2 BAO data vectors and their published covariance matrices.  Because BAO observables can be expressed in multiple, algebraically related forms, care must be taken to avoid double-counting correlated information.  Following DESI convention, a single, non-redundant BAO basis is adopted per analysis, and it is verified that no overlapping combinations are used.  This approach guarantees that any shift in parameter posteriors originates from the data basis or prior choice itself, not from hidden covariance overlap.

Three independent cases are analyzed
\begin{itemize}[leftmargin=*]
\item Scenario~A (mixed basis): $(D_V/r_d,\,D_M/D_H)$ with its full covariance, corresponding to $(\alpha_{\rm iso},\alpha_{\rm AP})$.  It separates the isotropic absolute-scale information from the anisotropic AP ratio and serves as our fiducial configuration.
\item Scenario~B (ratio-only): $(D_M/D_H)$ alone, a scale-free observable insensitive to $r_d$ and $H_0r_d$, used to isolate purely geometric constraints.  This case provides a diagnostic for potential over-interpretation of low-dimensional data.
\item Scenario~C (comparison): $(D_M/r_d,\,D_H/r_d)$, the conventional two-parameter basis adopted in earlier BAO analyses~\cite{Lee:2025rmg}, included for historical comparison.
\end{itemize}

The affine-invariant ensemble sampler \textsc{emcee}~\cite{Foreman-Mackey:2012any} is employed with 128 walkers, a burn-in phase of 4000 steps, and a production run of 12\,000 steps.  Convergence is verified through the integrated autocorrelation time and the stability of posterior means across independent chains.  The sampled parameters for the CPL model are $\boldsymbol{\theta}_{\wowaCDM} = (\Omegam, \,h\,r_d, \,\oo, \,\wa)$.  For the nested models,  $\boldsymbol{\theta}_{\wCDM} = (\Omegam0, \,h\,r_d, \,\oo)$ with $\wa=0$ and $\boldsymbol{\theta}_{\LCDM} = (\Omegam0, \,h\,r_d)$ with $\oo=-1$ and $\wa=0$ are adopted.  In this framework, the quantity $h\,r_d$ is treated as a single effective parameter encapsulating both the absolute Hubble scale and the BAO ruler.  The sound horizon $r_d$ itself is not independently computed from early-universe physics but absorbed into the sampling combination $h\,r_d$, which fully determines all BAO observables.  This treatment avoids introducing additional priors from CMB-based $r_d$ calibrations and ensures that the analysis remains purely BAO-driven.

The likelihood function is defined as
\begin{equation}
\mathcal{L}(\boldsymbol{\theta}|\mathbf{D})
\propto
\exp\!\left[-\tfrac{1}{2}\chi^2(\boldsymbol{\theta})\right],
\end{equation}
with
\begin{equation}
\chi^2(\boldsymbol{\theta})
=
(\mathbf{D}^{\mathrm{obs}}-\mathbf{D}^{\mathrm{th}}(\boldsymbol{\theta}))^{\!\top}
\mathbf{C}^{-1}
(\mathbf{D}^{\mathrm{obs}}-\mathbf{D}^{\mathrm{th}}(\boldsymbol{\theta})),
\end{equation}
where $\mathbf{C}$ is the covariance matrix provided by DESI for each observable combination.  

Flat (uninformative) priors are imposed on all sampled parameters to minimize bias:
\[
0.01 \leq \Omegam \leq 0.99, \qquad
-3.0 \leq \oo \leq 3.0, \qquad
-5.0 \leq \wa \leq 5.0.
\]
To assess robustness, two prior widths on $\wa$ are explored: the narrow range $[-2.5,2.5]$ and the wide range $[-5,5]$.  Comparing these runs isolates whether apparent deviations from $\LCDM$ reflect genuine data sensitivity or simply the expansion of parameter volume along the degeneracy ridge.

The median and central 68\% credible intervals of all parameters are reported, together with $\chi^2_{\min}$, reduced $\tilde{\chi}^2$, and information criteria (AIC, BIC).  Model-selection metrics are computed relative to $\LCDM$ using both the likelihood-ratio test (LRT) and the Savage–Dickey density ratio (SDDR) for nested models~\cite{Trotta:2008qt,Bridges:2008ta,Higson:2018cqj,Planck:2018vyg}.  
These complementary frequentist and Bayesian diagnostics quantify whether adding $(\oo,\,\wa)$ truly improves model fit or merely exploits prior volume.

All statistical comparisons are performed strictly within a given data basis, not across scenarios, to avoid conflating distinct likelihood definitions.  Ratio-only fits often yield $\tilde{\chi}^2_\nu\!\ll\!1$ because a single-observable dataset is over-parameterized;  such cases are interpreted as statistical overfitting rather than evidence for new physics.  
This methodology thus isolates the genuine information content of each DESI observable and disentangles physical constraints from basis- or prior-induced artifacts.

\section{Results}
\label{sec:results}

This section presents the cosmological parameter constraints obtained from the DESI~DR2 BAO likelihood under the three analysis scenarios defined in Sec.~\ref{sec:methodology}.  In all runs, the combined parameter $h\,r_d$ is sampled directly and $r_d$ is not computed from a CMB-based calibration or a fitting formula.  All theoretical predictions are therefore expressed in terms of $h\,r_d$ (and ratios such as $D_M/D_H$), ensuring that the analysis remains purely BAO-driven and free of external $r_d$ priors.  Unless otherwise noted, quoted values are posterior medians with central 68\% credible intervals, obtained from the MCMC setup described in Sec.~\ref{sec:methodology}.

\subsection*{Scenario A: Mixed basis $(D_V/r_d,\,D_M/D_H)$}
\label{subsec:SA}

\begin{figure}[htbp]
\centering
\includegraphics[width=0.8\textwidth]{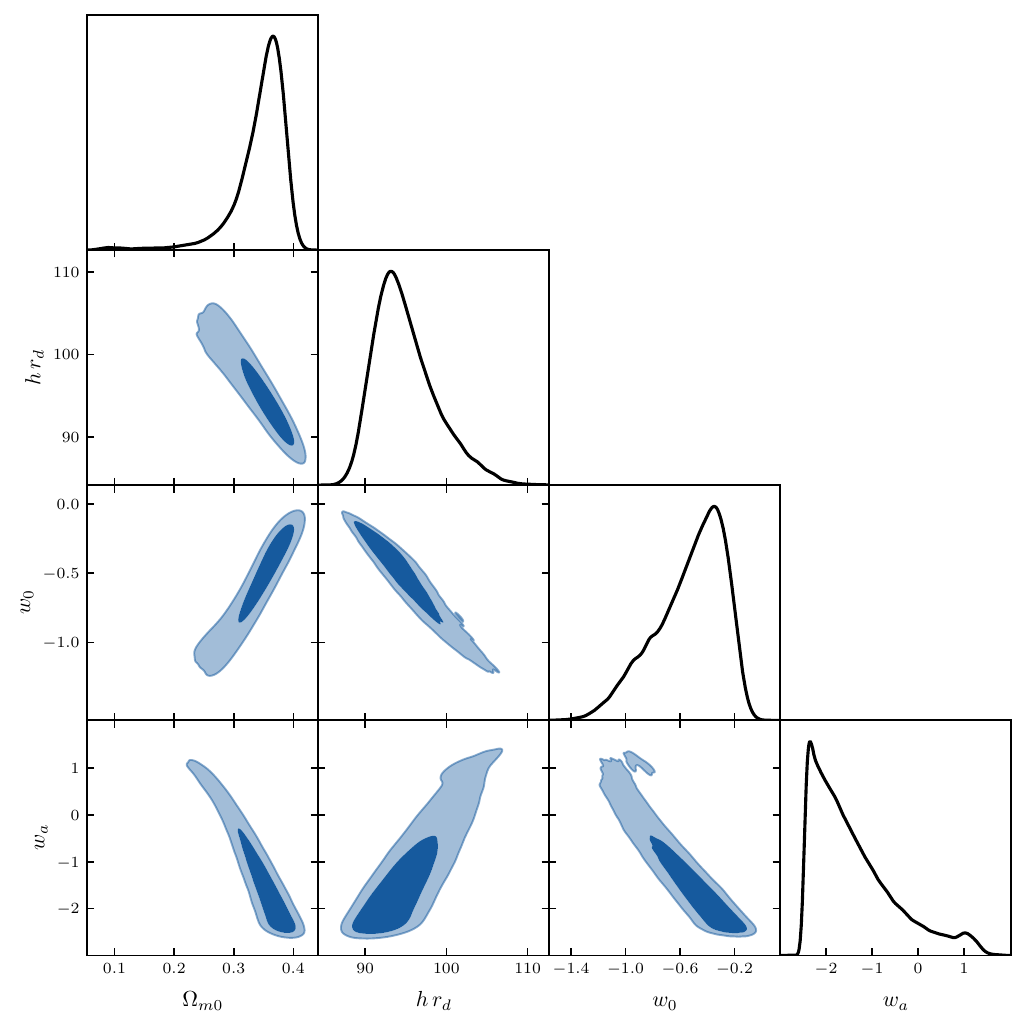}
\caption{Posterior distributions for $\Omegam$, $h r_d$, $\oo$, and $\wa$ in $\wowaCDM^{(25)}$ from Scenario A using $(D_V/r_d,\,D_M/D_H)$.  
Shaded regions indicate 68\% and 95\% credible contours.}
\label{fig:1}
\end{figure}

Figure~\ref{fig:1}, which was generated using the statistical plotting package \textsc{GetDist}~\cite{Lewis:2019xzd}, shows the marginalized posterior distributions for the $\wowaCDM^{(25)}$ run using the DESI reporting basis. In the following, $\wowaCDM^{(25)}$ denotes the CPL model analyzed with a narrow prior $\wa\in[-2.5,2.5]$, and $\wowaCDM^{(5)}$ denotes the corresponding case with a wide prior $\wa\in[-5.0,5.0]$. These two runs are used to assess how the choice of prior width influences the posterior distributions and the interpretation of DDE.
The contours reveal three characteristic features: 
(i) a pronounced anti-correlation between $(\oo,\,\wa)$ delineating the CPL degeneracy ridge, 
(ii)~a strong $\Omegam$–$h\,r_d$ anti-correlation from $D_V/r_d$, and 
(iii)~orthogonal constraints from $D_M/D_H$ that tighten the joint posteriors. Because the analysis samples the combined parameter $h\,r_d$ rather than computing $r_d$ from a CMB-based calibration, these results are fully BAO-driven and free from early-universe priors.

Table~\ref{tab:pars-fit-DVDMDH} summarizes the posterior constraints and fit quality for $\LCDM$, $\wCDM$, and two CPL variants with narrow ($\pm2.5$) and wide ($\pm5$) priors on $\wa$.  For the $\LCDM$ baseline, $\Omegam=0.295^{+0.010}_{-0.009}$ and $h\,r_d=101.8\pm0.8$ are obtained, giving $\chi^2_{\min}=10.02$ for ten degrees of freedom ($\tilde\chi^2\simeq1.00$).  
Allowing a constant-$\omega$ extension improves the fit only marginally, to $\chi^2_{\min}=9.51$ at $\oo=-0.93\pm0.09$, a shift of $\Delta\chi^2\simeq0.5$ that is statistically insignificant ($p_{\rm LRT}\!\approx\!0.48$).  
Introducing an evolving EoS through the CPL parametrization reduces the minimum $\chi^2$ further—to $6.25$ in the narrow-prior run and $5.16$ in the wide-prior run—yet these changes correspond to $\Delta\chi^2\lesssim5$ for two extra parameters, yielding $p_{\rm LRT}\simeq0.1$--0.15.  
The modest decrease in $\chi^2$ is therefore offset by the AIC/BIC penalties: $\Delta{\rm AIC}=0.23$--$0.86$ and $\Delta{\rm BIC}=1.2$--$1.9$, indicating that all models remain statistically indistinguishable.

The two CPL cases illustrate how prior width affects parameter excursions along the degeneracy ridge.  
With $\wa\!\in[-2.5,2.5]$, the fit gives $\oo=-0.48^{+0.20}_{-0.33}$ and $\wa=-1.63^{+1.15}_{-0.63}$ at $\Omegam=0.36^{+0.03}_{-0.04}$ and $h\,r_d=94.4^{+4.5}_{-3.0}$.  
Expanding the prior to $\wa\!\in[-5,5]$ drives the posterior toward $\oo=-0.13^{+0.33}_{-0.44}$ and $\wa=-2.79^{+1.46}_{-1.10}$, together with a slightly higher $\Omegam=0.40^{+0.04}_{-0.05}$ and a lower $h\,r_d=90.3^{+5.4}_{-3.5}$.  
These shifts trace the known anti-correlation between $(\oo,\,\wa)$ and between $(\Omegam,\,h\,r_d)$ rather than signaling any genuine DE evolution.  Although both CPL fits yield smaller reduced $\tilde{\chi}^2$ values ($0.78$ and $0.65$) compared to unity, this improvement reflects the enlarged parameter volume rather than additional information content.  
Consequently, the apparent evolution inferred from broader priors is best interpreted as movement along the CPL degeneracy ridge rather than as evidence for new physics.

\begin{table*}[t]
\centering
\caption{Posterior constraints and fit quality for $\Lambda$CDM, $\wCDM$, and $\wowaCDM$ using DESI DR2 BAO data in the $(D_V/r_d,\,D_M/D_H)$ basis.
Reported values are medians with central 68\% credible intervals.
Akaike and BIC weights ($W_{\rm AIC},W_{\rm BIC}$) are normalized over the four models.
The superscripts $(25)$ and $(5)$ denote the narrow and wide prior ranges on $\wa$,
corresponding to $\wa\in[-2.5,2.5]$ and $\wa\in[-5.0,5.0]$, respectively.}
\label{tab:pars-fit-DVDMDH}
\begin{ruledtabular}
\begin{tabular}{lcccccccccc}
\toprule
Model & $\oo$ & $\wa$ & $\Omegam$ & $h r_d$ & $\chi^2_{\min}$ (dof) & $\tilde\chi^2$ & AIC & BIC & $W_{\rm AIC}$ & $W_{\rm BIC}$ \\
\hline
\addlinespace[0.3em]
$\Lambda$CDM & $-1$ & $0$ & $0.2951^{+0.0096}_{-0.0093}$ & $101.78^{+0.84}_{-0.83}$ & $10.022\ (10)$ & $1.002$ & $14.022$ & $14.992$ & $0.26$ & $0.27$ \\
\addlinespace[0.3em]
$\wCDM$ & $-0.925^{+0.090}_{-0.094}$ & $0$ & $0.2960^{+0.0101}_{-0.0098}$ & $100.03^{+2.37}_{-2.17}$ & $9.512\ (9)$ & $1.057$ & $15.512$ & $16.967$ & $0.14$ & $0.12$ \\
\addlinespace[0.3em]
$\wowaCDM^{(25)}$ & $-0.482^{+0.202}_{-0.328}$ & $-1.63^{+1.15}_{-0.63}$ & $0.355^{+0.026}_{-0.041}$ & $94.36^{+4.50}_{-3.03}$ & $6.251\ (8)$ & $0.781$ & $14.251$ & $16.191$ & $0.31$ & $0.30$ \\
\addlinespace[0.3em]
$\wowaCDM^{(5)}$ & $-0.131^{+0.327}_{-0.443}$ & $-2.79^{+1.46}_{-1.10}$ & $0.395^{+0.037}_{-0.051}$ & $90.28^{+5.36}_{-3.48}$ & $5.160\ (8)$ & $0.645$ & $13.160$ & $15.100$ & $0.29$ & $0.31$ \\
\bottomrule
\end{tabular}
\end{ruledtabular}
\end{table*}

Model-selection statistics in Table~\ref{tab:msel-DVDMDH} confirm that $\Lambda$CDM remains moderately preferred.  
Bayes factors $B_{01}\simeq8$–9 correspond to “substantial” evidence on Jeffreys’ scale, while $\Delta$AIC and $\Delta$BIC below 2 indicate that the CPL extensions are statistically indistinguishable from $\Lambda$CDM.  

\begin{table}[h]
\centering
\caption{Model-selection metrics relative to $\Lambda$CDM in the $(D_V/r_d,\,D_M/D_H)$ basis. 
$T_k=\chi^2_{\Lambda{\rm CDM}}-\chi^2_{\rm ext}$ with $k$ extra parameters; 
$p_{\rm LRT}$ is the $\chi^2_k$ tail probability. 
$B_{01}>1$ favors $\Lambda$CDM. Akaike and BIC weights are normalized across the four models.}
\label{tab:msel-DVDMDH}
\begin{ruledtabular}
\begin{tabular}{lcccccc}
\toprule
Extension & $T_k$ (dof) & $p_{\rm LRT}$ & $B_{01}$ & $W_{\rm AIC}$ & $W_{\rm BIC}$ & Notes \\
\hline
\addlinespace[0.3em]
$\wCDM$ & $0.510$ (1) & $0.475$ & $5.17$ & $0.14$ & $0.12$ & substantial support for $\Lambda$CDM \\
\addlinespace[0.3em]
$\wowaCDM^{(25)}$ & $3.771$ (2) & $0.152$ & $9.35$ & $0.31$ & $0.30$ & weak evidence for $\Lambda$CDM  \\
\addlinespace[0.3em]
$\wowaCDM^{(5)}$ & $4.862$ (2) & $0.088$ & $7.96$ & $0.29$ & $0.31$ & weak evidence for $\Lambda$CDM  \\
\bottomrule
\end{tabular}
\end{ruledtabular}
\end{table}

\paragraph*{Degeneracy structure and pivoted EoS.}
Posterior covariance matrices (Table~\ref{CVCCofDVDMDH}) show a persistent strong anti-correlation $\rho_{\oo\wa}\simeq -0.95$, defining a near one-dimensional likelihood ridge.  
Using Eq.~\eqref{wpp}, the pivoted equation of state is computed,
\begin{align}
\wpp = \oo + (1-a_p) \wa , \qquad a_p = 1+\frac{\mathrm{Cov}(\oo,\,\wa)}{\mathrm{Var}(\wa)}  \label{wpp2} ,
\end{align}
and obtain
\begin{align}
\wowaCDM^{(25)}:\; & a_p=0.739,\; w_p=-0.907\pm0.094 , \nonumber\\
\wowaCDM^{(5)}:\; & a_p=0.729,\; w_p=-0.886\pm0.094 .
\end{align}
Including the propagated uncertainty from the MCMC posterior covariance, the pivot scale is found to be
$a_p = 0.74 \pm 0.02$ for the narrow-prior case and $a_p = 0.73 \pm 0.02$ for the wide-prior case.  Both yield consistent pivot values $\wpp\simeq-0.9$ at $z_p\simeq0.35$, fully consistent with a cosmological constant within $1\sigma$.  Hence, the data constrain a single effective EoS combination $\wpp=\omega(a_p)$, while individual $(\oo,\,\wa)$ values slide along the degeneracy ridge without altering $\wpp$.  This behavior reproduces the well-known $\oo$–$\wa$ correlation geometry reported in earlier analyses~\cite{Szydlowski:2006pz,Gong:2013bn,Shlivko:2024llw}.

\begin{table}[htbp]
\centering
\caption{Posterior covariance and correlation matrices of cosmological parameters for each model in the $(D_V/r_d,\,D_M/D_H)$ basis. 
Correlation coefficients are dimensionless and highlight parameter degeneracies more clearly.}
\begin{ruledtabular}
\begin{tabular}{lccc}
\toprule
Model & Parameters & Cov & $\rho$ \\
\hline
\addlinespace[0.3em]
$\LCDM$ & [$\Omegam$, $h r_d$] &
$\begin{pmatrix}
0.00009038 & -0.00753666 \\
-0.00753666 & 0.7081198
\end{pmatrix}$ &
$\begin{pmatrix}
1.000 & -0.942 \\
-0.942 & 1.000
\end{pmatrix}$ \\[2ex]

$\wCDM$ & [$\Omegam$, $h r_d$, $\oo$] &
$\begin{pmatrix}
0.00010096 & -0.01125375 & 0.00017282 \\
-0.01125375 & 5.282848 & -0.200982 \\
0.00017282 & -0.200982 & 0.00868835
\end{pmatrix}$ &
$\begin{pmatrix}
1.000 & -0.487 & 0.185 \\
-0.487 & 1.000 & -0.938 \\
0.185 & -0.938 & 1.000
\end{pmatrix}$ \\[2ex]

$\wowaCDM^{(25)}$ & [$\Omegam$, $h r_d$, $\oo$, $\wa$] &
$\begin{pmatrix}
0.00201859 & -0.1492989 & 0.00959250 & -0.03683249 \\
-0.1492989 & 15.01968 & -0.9486064 & 2.982906 \\
0.00959250 & -0.9486064 & 0.06530136 & -0.2167885 \\
-0.03683249 & 2.982906 & -0.2167885 & 0.8321526
\end{pmatrix}$ &
$\begin{pmatrix}
1.000 & -0.857 & 0.836 & -0.899 \\
-0.857 & 1.000 & -0.958 & 0.844 \\
0.836 & -0.958 & 1.000 & -0.930 \\
-0.899 & 0.844 & -0.930 & 1.000
\end{pmatrix}$ \\[2ex]

$\wowaCDM^{(5)}$ & [$\Omegam$, $h r_d$, $\oo$, $\wa$] &
$\begin{pmatrix}
0.00324725 & -0.2185163 & 0.01749871 & -0.06590931 \\
-0.2185163 & 19.71837 & -1.565186 & 5.217232 \\
0.01749871 & -1.565186 & 0.1337685 & -0.4617110 \\
-0.06590931 & 5.217232 & -0.4617110 & 1.705335
\end{pmatrix}$ &
$\begin{pmatrix}
1.000 & -0.864 & 0.840 & -0.886 \\
-0.864 & 1.000 & -0.964 & 0.900 \\
0.840 & -0.964 & 1.000 & -0.967 \\
-0.886 & 0.900 & -0.967 & 1.000
\end{pmatrix}$ \\
\bottomrule
\end{tabular}
\end{ruledtabular}
\label{CVCCofDVDMDH}
\end{table}

\paragraph*{Interpretation.}
The mixed basis provides balanced constraints by combining absolute- and ratio-based observables.  All models give excellent fits ($\tilde\chi^2\!\approx\!0.7$–1.0), and none show significant evidence for DDE.  The apparent evolution suggested by the CPL parameters arises from prior width and degeneracy geometry, not from any new physical signal.  The pivoted EoS $\wpp\simeq-0.9\pm0.1$ remains statistically indistinguishable from $\omega=-1$, underscoring that DESI~DR2 BAO data alone contain no preference for DDE once parameter correlations are properly accounted for.

\begin{figure*}[t]
  \centering
  \includegraphics[width=0.95\textwidth]{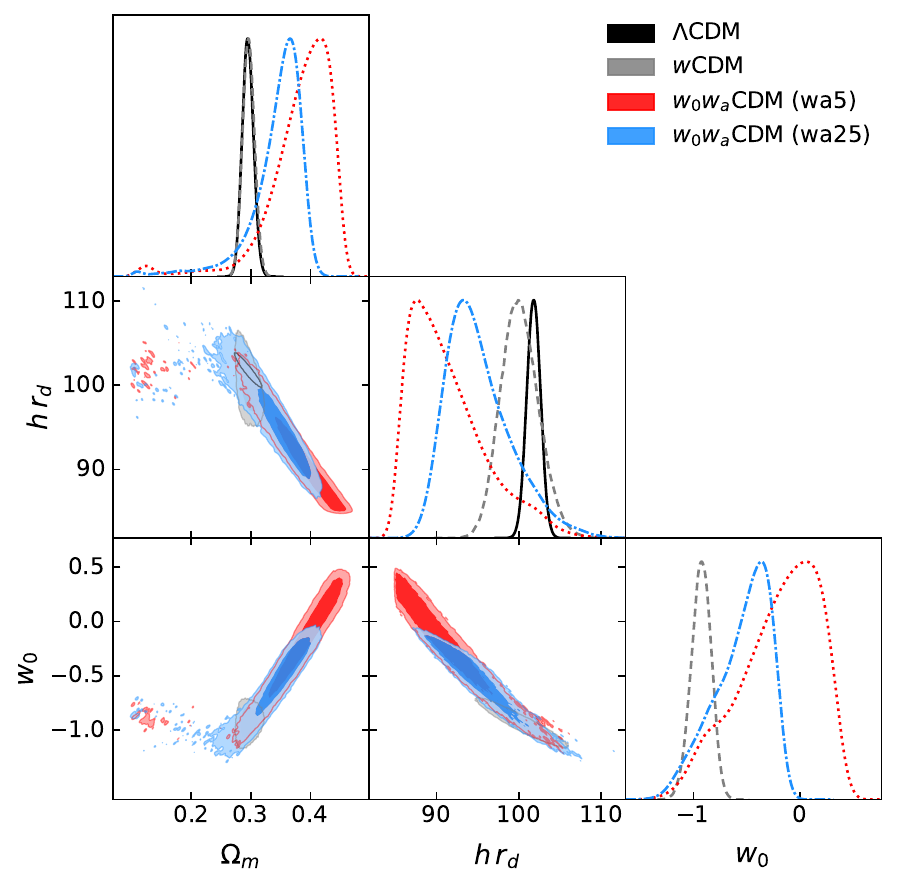}
  \caption{Summary of Scenario~A constraints obtained from the mixed DESI basis
$(D_V/r_d,\,D_M/D_H)$.
The figure overlays the marginalized posteriors in the
$(\Omegam,\,h r_d,\,\oo)$ parameter subspace for four cosmological models:
$\LCDM$ (black solid),
$\wCDM$ (gray dashed),
$\wowaCDM$ with $\wa\in[-5,5]$ (red dotted),
and $\wowaCDM$ with $\wa\in[-2.5,2.5]$ (blue dash-dotted).
}
\label{fig:2}
\end{figure*}

\begin{figure}[t]
  \centering
  \includegraphics[width=0.55\columnwidth]{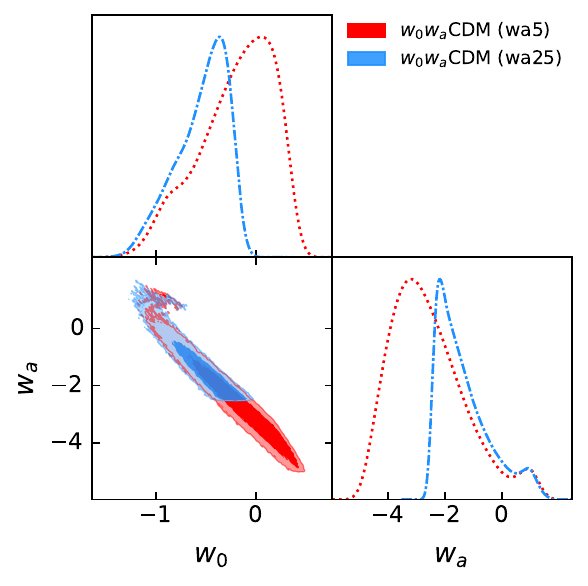}
\caption{
Prior-projection effect in Scenario~A illustrated in the $(\oo,\,\wa)$ plane
for the $\wowaCDM$ model.
Posteriors obtained with a flat prior $\wa\in[-5,5]$ are shown as
red dotted contours,
while those obtained with $\wa\in[-2.5,2.5]$ are shown as
blue dash-dotted contours.
The apparent shift reflects truncation of the posterior volume along the
intrinsic BAO degeneracy direction.
}
  \label{fig:3}
\end{figure}

Figures~\ref{fig:2} and \ref{fig:3} provide a compact visual summary of the numerical results reported in Tables~\ref{tab:pars-fit-DVDMDH} and \ref{CVCCofDVDMDH} for Scenario~A. Even with the inclusion of an absolute-distance observable ($D_V/r_d$), the mixed basis $(D_V/r_d,\,D_M/D_H)$ admits a highly anisotropic posterior geometry, allowing substantial parameter motion along the CPL degeneracy direction with little change in fit quality. Figure~\ref{fig:3} makes this mechanism explicit: tightening the flat prior range of $\wa$ mainly truncates the posterior volume along the nearly flat likelihood ridge, thereby shifting the apparent $(\oo,\wa)$ preference without introducing new information.

Both figures may display small isolated contour fragments and fine substructures. These are caused by local variations in sampling density, combined with the strongly elongated BAO degeneracy and the very large effective chain size; they do not indicate distinct physical modes and do not affect the global degeneracy directions or the comparative conclusions emphasized here.

\subsection*{Scenario B: Ratio-only $(D_M/D_H)$}
\label{subsec:DMDH}

\begin{figure}[htbp]
\centering
\includegraphics[width=0.8\textwidth]{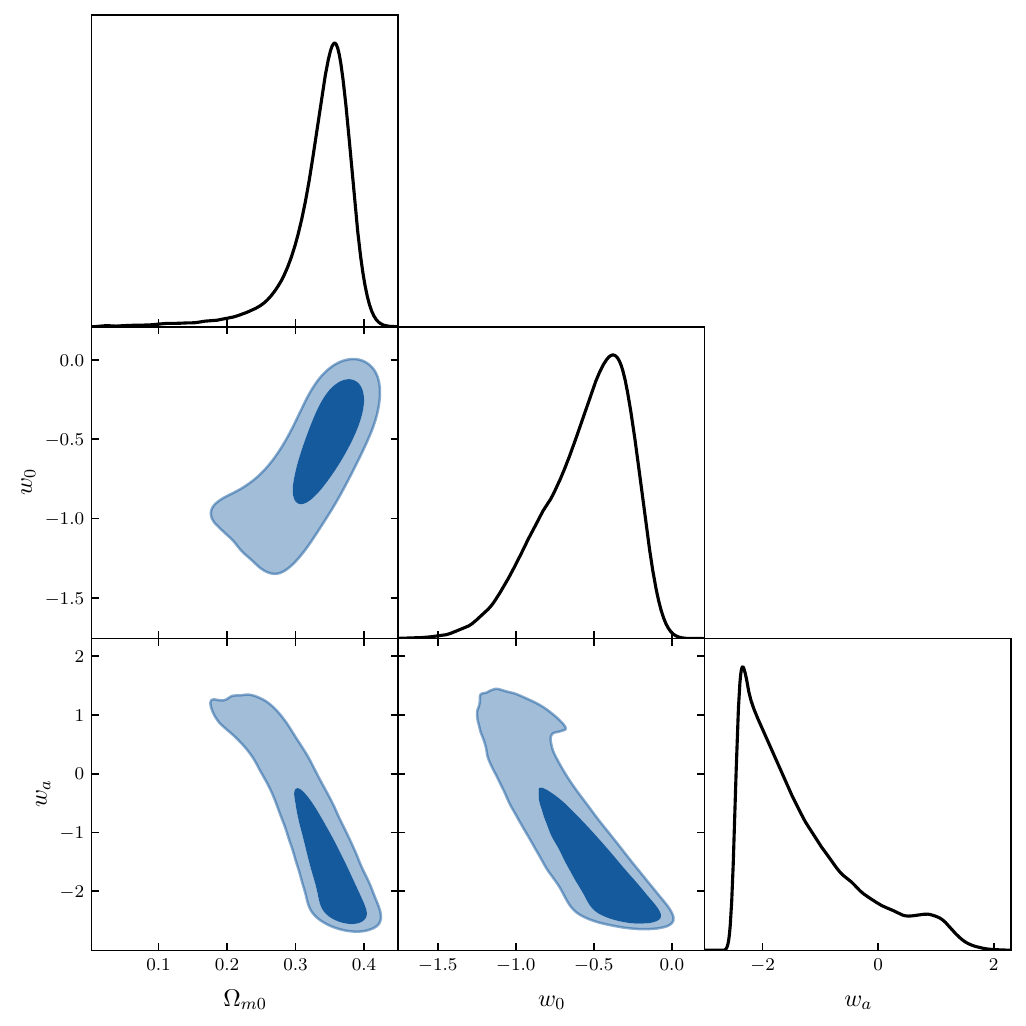}
\caption{Posterior distributions for $\Omegam$, $\oo$, and $\wa$ in $\wowaCDM^{(25)}$ from Scenario B using $D_M/D_H$ only.  The extremely low reduced $\tilde{\chi}^2$ in the wide-prior run ($\tilde{\chi}^2\!\simeq\!0.42$) indicates mild overfitting of a low-dimensional dataset, not genuine physical improvement.}
\label{fig:4}
\end{figure}

Figure~\ref{fig:4} displays the marginalized posteriors from the $D_M/D_H$-only likelihood.  Because this observable is scale-free, the contours trace an elongated degeneracy ridge with very weak constraints on $\Omegam$ and a strong anti-correlation between $\oo$ and $\wa$.  Since $D_M/D_H$ depends only on the dimensionless expansion rate $E(z)$, it constrains the shape of $H(z)$ but carries no information about the absolute distance scale or $h\,r_d$.  The resulting posterior therefore reflects pure geometric ratios rather than anchored distances.  Widening the $\wa$ prior substantially shifts the posterior mean while leaving $\chi^2_{\min}$ almost unchanged, illustrating the susceptibility of ratio-only analyses to prior-volume effects.

Table~\ref{tab:pars-fit-DMDH} summarizes the parameter constraints and fit quality for the four models.  For the $\LCDM$ baseline, $\Omegam=0.300^{+0.018}_{-0.017}$ and $\chi^2_{\min}=6.04$ (five degrees of freedom), corresponding to $\tilde{\chi}^2\simeq1.21$.  Allowing $\oo$ to vary in $\wCDM$ reduces $\chi^2_{\min}$ only slightly to $5.43$, yielding $p_{\rm LRT}\simeq0.44$.  Introducing a time-dependent EoS via the CPL form lowers $\chi^2_{\min}$ to $3.06$ for $\wowaCDM^{(25)}$ and $1.25$ for $\wowaCDM^{(5)}$, but the $\Delta\chi^2\simeq2$--$5$ improvement is statistically insignificant once the two extra parameters are penalized.  Indeed, AIC and BIC weights remain dominated by $\LCDM$, with $W_{\rm AIC,BIC}\!\approx\!0.5$ in the wide-prior run and $\approx0.1$ in the narrow-prior run, both indicating weak or substantial support for $\LCDM$.

\begin{table*}[t]
\centering
\caption{Posterior constraints and fit quality for $\Lambda$CDM, $\wCDM$, and $\wowaCDM$ using DESI DR2 BAO data in the $(D_M/D_H)$ basis.  
Reported values are medians with central 68\% credible intervals.  
The two CPL runs differ only by their prior ranges, denoted $\wowaCDM^{(25)}$ and $\wowaCDM^{(5)}$.  
Akaike and BIC weights ($W_{\rm AIC},W_{\rm BIC}$) are normalized across the four models.}
\label{tab:pars-fit-DMDH}
\begin{ruledtabular}
\begin{tabular}{lccccccccc}
\toprule
Model & $\oo$ & $\wa$ & $\Omegam$ & $\chi^2_{\min}$ (dof) & $\tilde\chi^2$ & AIC & BIC & $W_{\rm AIC}$ & $W_{\rm BIC}$ \\
\hline
\addlinespace[0.3em]
$\Lambda$CDM & $-1$ & $0$ & $0.300^{+0.018}_{-0.017}$ & $6.036\ (5)$ & $1.207$ & $8.036$ & $7.828$ & $0.26$ & $0.22$ \\
\addlinespace[0.3em]
$\wCDM$ & $-0.898^{+0.128}_{-0.145}$ & $0$ & $0.293^{+0.020}_{-0.021}$ & $5.431\ (4)$ & $1.358$ & $9.431$ & $9.015$ & $0.11$ & $0.12$ \\
\addlinespace[0.3em]
$\wowaCDM^{(25)}$ & $-0.504^{+0.224}_{-0.338}$ & $-1.61^{+1.33}_{-0.65}$ & $0.346^{+0.029}_{-0.046}$ & $3.055\ (3)$ & $1.018$ & $9.055$ & $8.431$ & $0.11$ & $0.13$ \\
\addlinespace[0.3em]
$\wowaCDM^{(5)}$ & $+0.035^{+0.319}_{-0.487}$ & $-3.51^{+1.70}_{-1.06}$ & $0.405^{+0.037}_{-0.053}$ & $1.254\ (3)$ & $0.418$ & $7.254$ & $6.629$ & $0.52$ & $0.53$ \\
\bottomrule
\end{tabular}
\end{ruledtabular}
\end{table*}

The effect of enlarging the $\wa$ prior is clearly visible in the parameter shifts.  With $\wa\in[-2.5,2.5]$, $\oo=-0.50^{+0.22}_{-0.34}$, $\wa=-1.61^{+1.33}_{-0.65}$, and $\Omegam=0.35^{+0.03}_{-0.05}$ are obtained.  
Expanding to $\wa\in[-5,5]$ displaces the posterior along the degeneracy ridge toward $\oo=+0.04^{+0.32}_{-0.49}$ and $\wa=-3.51^{+1.70}_{-1.06}$, accompanied by a higher $\Omegam=0.41^{+0.04}_{-0.05}$.  The near-constant $\chi^2_{\min}$ across these runs shows that the shift arises from prior volume rather than new information.  The extremely low reduced $\tilde{\chi}^2\simeq0.42$ for the wide-prior case reflects mild overfitting caused by introducing two extra parameters to fit only one independent observable.

\begin{table}[h]
\centering
\caption{Model-selection metrics relative to $\LCDM$ in the $(D_M/D_H)$ basis.  $T_k=\chi^2_{\LCDM}-\chi^2_{\rm ext}$ with $k$ extra parameters.  $B_{01}$ denotes the Bayes factor from SDDR.  Akaike/BIC weights are normalized across all models.}
\label{tab:msel-DMDH}
\begin{ruledtabular}
\begin{tabular}{lcccccc}
\toprule
Extension & $T_k$ (dof) & $p_{\rm LRT}$ & $B_{01}$ & $W_{\rm AIC}$ & $W_{\rm BIC}$ & Notes \\
\hline
\addlinespace[0.3em]
$\wCDM$ & $0.605$ (1) & $0.44$ & $3.54$ & $0.11$ & $0.12$ & substantial support for $\Lambda$CDM \\
\addlinespace[0.3em]
$\wowaCDM^{(25)}$ & $2.981$ (2) & $0.22$ & $7.39$ & $0.11$ & $0.13$ & weak evidence for $\Lambda$CDM  \\
\addlinespace[0.3em]
$\wowaCDM^{(5)}$ & $4.782$ (2) & $0.09$ & $4.03$ & $0.52$ & $0.53$ & weak evidence for $\Lambda$CDM \\
\bottomrule
\end{tabular}
\end{ruledtabular}
\end{table}

\begin{table}[htbp]
\centering
\caption{Posterior covariance and correlation matrices of cosmological parameters for each model in the $(D_M/D_H)$-only basis. 
Correlation coefficients are dimensionless and highlight parameter degeneracies more clearly.}
\begin{ruledtabular}
\begin{tabular}{lccc}
\toprule
Model & Parameters & Cov & $\rho$ \\
\hline
\addlinespace[0.3em]
$\Lambda$CDM & [$\Omegam$] &
$\begin{pmatrix}
0.0003129
\end{pmatrix}$ &
$\begin{pmatrix}
1.000
\end{pmatrix}$ \\[2ex]

$\wCDM$ & [$\Omegam$, $\oo$] &
$\begin{pmatrix}
0.0004521 & -0.001283 \\
-0.001283 & 0.019511
\end{pmatrix}$ &
$\begin{pmatrix}
1.000 & -0.432 \\
-0.432 & 1.000
\end{pmatrix}$ \\[2ex]

$\wowaCDM^{(25)}$ & [$\Omegam$, $\oo$, $\wa$] &
$\begin{pmatrix}
0.002541 & 0.009499 & -0.042270 \\
0.009499 & 0.076104 & -0.239076 \\
-0.042270 & -0.239076 & 0.973030
\end{pmatrix}$ &
$\begin{pmatrix}
1.000 & 0.683 & -0.850 \\
0.683 & 1.000 & -0.879 \\
-0.850 & -0.879 & 1.000
\end{pmatrix}$ \\[2ex]

$\wowaCDM^{(5)}$ & [$\Omegam$, $\oo$, $\wa$] &
$\begin{pmatrix}
0.002565 & 0.017797 & -0.063438 \\
0.017797 & 0.160996 & -0.527214 \\
-0.063438 & -0.527214 & 1.926180
\end{pmatrix}$ &
$\begin{pmatrix}
1.000 & 0.876 & -0.902 \\
0.876 & 1.000 & -0.947 \\
-0.902 & -0.947 & 1.000
\end{pmatrix}$ \\
\bottomrule
\end{tabular}
\end{ruledtabular}
\label{CVCCofDMDH}
\end{table}

Posterior covariances in Table~\ref{CVCCofDMDH} confirm that $\rho_{\oo\,\wa}\simeq-0.90$, consistent with Scenario~A and previous studies~\cite{Szydlowski:2006pz,Gong:2013bn,Shlivko:2024llw}.  Computing the pivoted equation of state yields $\wpp=-0.91\pm0.13$ at $z_p\simeq0.34$, statistically identical to the mixed-basis result ($\wpp\simeq-0.9\pm0.1$).  
Thus, the ratio-only likelihood preserves the same degeneracy geometry but lacks the absolute-scale anchor required to distinguish between constant and DDE.

\paragraph*{Interpretation.}
The $D_M/D_H$ observable serves primarily as a geometric consistency test.  While it constrains the AP ratio effectively, its scale-free nature prevents it from breaking degeneracies among $(\Omegam,\,h\,r_d,\,\oo,\,\wa)$.  
Consequently, even large apparent shifts in $(\oo,\,\wa)$ can occur without any significant $\Delta\chi^2$.  
This scenario highlights the importance of including an absolute-distance observable—such as $D_V/r_d$—to anchor the BAO scale and avoid misinterpreting prior-induced parameter drifts as DDE evidence.

\subsection*{Scenario C (DESI basis): $(D_M/r_d,\,D_H/r_d)$}
\label{subsec:DMrdDHrd}

Using the conventional $(D_M/r_d,\,D_H/r_d)$ basis—historically the most common choice in BAO analyses—it is tested whether the inferred CPL parameters depend on prior width or data basis.  As in Scenarios~A and~B, $r_d$ is not computed from CMB calibration but absorbed into the sampled combination $h\,r_d$, keeping the analysis BAO-only.  This configuration allows direct comparison with previous DESI-style analyses and isolates any basis-induced effects.

Widening the prior from $[-2.5,2.5]$ to $[-5,5]$ primarily shifts the posterior along the well-known $\oo$–$\wa$ degeneracy ridge without improving the fit quality.  
The detailed constraints are:
\begin{align}
\wowaCDM^{(25)}:&\ 
\Omegam=0.343^{+0.038}_{-0.030},\quad h r_d=94.98^{+3.88}_{-3.67}, \nonumber \\
&\ \oo=-0.53^{+0.25}_{-0.28},\quad \wa=-1.38^{+0.92}_{-0.88},\quad 
\chi^2_{\min}=6.193\ (\tilde{\chi}^2=0.774), \nonumber \\
\wowaCDM^{(5)}:&\ 
\Omegam=0.389^{+0.043}_{-0.042},\quad h r_d=90.99^{+4.26}_{-4.23}, \label{wowacdmDMrdDHrd} \\
&\ \oo=-0.17^{+0.37}_{-0.38},\quad \wa=-2.66^{+1.25}_{-1.24},\quad 
\chi^2_{\min}=5.108\ (\tilde{\chi}^2=0.639), \nonumber \\[0.2em]
\wowaCDM^{(\mathrm{DESI})}:&\ 
\Omegam=0.361^{+0.035}_{-0.034},\quad h r_d=93.53^{+3.79}_{-3.68}, \nonumber \\
&\ \oo=-0.41^{+0.27}_{-0.29},\quad \wa=-1.81^{+0.94}_{-0.91},\quad 
\wa<-1.54\ (68\%~\mathrm{upper~limit}), \quad
\chi^2_{\min}=5.706\ (\tilde{\chi}^2=0.713) . \nonumber 
\end{align}
The $\wowaCDM^{(\mathrm{DESI})}$ case corresponds to a run with $\oo\in[-3,0]$ and $\wa\in[-5,5]$,  
where the one-sided upper limit on $\wa$ follows the same statistical convention as Table~V of the DESI DR2 paper.  
The decrease in $\chi^2_{\min}$ from $6.19$ to $5.11$ corresponds to $\Delta\chi^2\simeq1.1$ for identical degrees of freedom—statistically insignificant.  However, the apparent changes in $(\oo,\,\wa)$, from $(-0.53,-1.38)$ to $(-0.17,-2.66)$, are sizable, again illustrating how prior expansion moves the posterior along the degeneracy ridge without altering the goodness of fit.

The pivoted equation of state remains nearly invariant:
\begin{align}
\wpp^{(25)} &= -0.888 \pm 0.099\quad (a_p=0.751,\ z_p\simeq0.33), \nonumber \\
\wpp^{(5)}  &= -0.909 \pm 0.083\quad (a_p=0.721,\ z_p\simeq0.39), \label{wppCaseC} \\
\wpp^{(\mathrm{DESI})}  &= -0.897 \pm 0.084\quad (a_p=0.733,\ z_p\simeq0.36) \nonumber ,
\end{align}
all consistent with $w=-1$ at the $1\sigma$ level.  Thus, even though $(\oo,\,\wa)$ shift substantially with prior width,  the DESI basis reproduces the same pivoted value $\wpp\simeq-0.9\pm0.1$ obtained in Scenarios~A and~B, reaffirming that the physically constrained quantity is $\wpp$ rather than the individual $(\oo,\,\wa)$.

\paragraph*{Interpretation.}
A superficial comparison might suggest that the observed shifts in $(\oo,\wa)$ arise from the choice of BAO basis.  However, the results show that basis choice affects only the covariance geometry, while the prior width on $\wa$ dictates the apparent location along the degeneracy ridge.  For identical $\wa$ priors, the mixed $(D_V/r_d,\,D_M/D_H)$ and DESI $(D_M/r_d,\,D_H/r_d)$ analyses give consistent results: $(\oo,\,\wa)=(-0.48,-1.63)$ versus $(-0.53,-1.38)$ for $\wa\in[-2.5,2.5]$,  and $(-0.13,-2.79)$ versus $(-0.17,-2.66)$ for $\wa\in[-5,5]$.  In all cases, the pivoted equation of state remains stable at $\wpp\simeq-0.9$ and $z_p\simeq0.33$–0.39, confirming that the underlying physics is invariant to the adopted basis.

\paragraph*{Caution on external priors.}
Combining BAO data with external probes such as the CMB or supernovae imposes strong priors on $\Omegam$, $h$, or $r_d$, which can artificially compress the $(\oo,\,\wa)$ posterior along the degeneracy ridge.  This often drives $\wa$ toward negative values and produces apparent DDE signals that arise from the external prior rather than the data itself ~\cite{Sakr:2025daj,Jiang:2025ylr,Huang:2025som,Lee:2025kbn,Toomey:2025xyo}.  The BAO-only result $\wpp\simeq-0.9\pm0.1$ therefore serves as a baseline against which the influence of external priors can be quantified

\paragraph*{Limitation of the CPL parametrization.}
The CPL form $\omega(a)=\oo+\wa(1-a)$ offers computational simplicity but restricted physical flexibility.  Because its linear evolution cannot describe non-monotonic or rapidly varying $\omega(a)$ histories,  and because $\oo$ and $\wa$ remain strongly anti-correlated, current data effectively constrain only the pivoted value $\wpp=\omega(a_p)$.  Consequently, frequent findings of $\wa<0$ in CPL-based studies largely reflect degeneracy geometry and prior choice rather than true DE dynamics.  
Future tests of DDE should employ non-parametric or physically motivated extensions beyond the CPL ansatz.

\paragraph*{Summary of results.}
Across all scenarios, the key findings are:
\begin{itemize}[leftmargin=*]
\item Apparent shifts in $(\oo,\wa)$ reflect sampling along a nearly flat degeneracy ridge; $\LCDM$ remains within $1\sigma$.
\item The pivoted EoS is stable, $\wpp\simeq-0.9\pm0.1$, independent of basis or prior.
\item Model-selection metrics ($\Delta$AIC, $\Delta$BIC $\lesssim$ 2; $B_{01}>3$) indicate no statistically significant improvement over $\LCDM$.
\item Ratio-only data tend to overfit and should not be interpreted as evidence for evolving $\omega(a)$.
\end{itemize}
Together, these results demonstrate that the DESI~DR2 BAO likelihood—across all basis and prior choices—is fully consistent with a cosmological constant once degeneracy geometry and prior sensitivity are properly accounted for.

\section{Conclusion}
\label{sec:conclusion}

Across all BAO bases examined—$(D_V/r_d,\,D_M/D_H)$, $(D_M/D_H)$, and $(D_M/r_d,\,D_H/r_d)$—the cosmological constraints derived from the DESI~DR2 likelihood exhibit a common geometric structure dominated by the well-known degeneracy between $\oo$ and $\wa$.  The principal degeneracy direction in the $(\oo,\,\wa)$ plane remains nearly identical across bases, corresponding to the smallest-eigenvalue mode of the Fisher matrix.  Apparent shifts in $(\oo,\,\wa)$ therefore trace the same one-dimensional ridge rather than reflecting genuine dark-energy evolution.  These directions correspond to parameter combinations that leave the observed distance ratios $(D_M/r_d,\,D_H/r_d,\,D_M/D_H)$ nearly invariant, and thus arise from the geometry of the BAO likelihood itself rather than from new physics.

Because $D_M/D_H$ is a scale-free quantity, analyses relying solely on this ratio suffer from elongated degeneracies and extremely low reduced $\tilde{\chi}^2$ values, indicating mild overfitting of a low-dimensional dataset.  The inclusion of an absolute-scale observable such as $D_V/r_d$ is therefore essential to anchor the BAO ruler and recover physically interpretable constraints.  When both observables are used consistently within a single non-redundant basis, all scenarios yield compatible posteriors: for narrow and wide priors on $\wa$, the pivoted equation of state remains stable at $\wpp=\omega(a_p)\simeq-0.9\pm0.1$ near $z_p\simeq0.34$.  This quantity, rather than the individual $(\oo,\,\wa)$ values, represents the truly constrained combination of parameters.

Statistical comparisons further support this interpretation.  Differences in $\chi^2_{\min}$ between models are small ($\Delta\chi^2\lesssim5$), and the corresponding improvements vanish once penalized by AIC and BIC.  Bayes factors computed via the Savage–Dickey density ratio yield $B_{01}\simeq4$–9, corresponding to moderate support for $\LCDM$ on Jeffreys’ scale.  Taken together, frequentist and Bayesian metrics converge on the conclusion that the CPL extension, while flexible, offers no statistically significant improvement over $\LCDM$.  Low $\tilde{\chi}^2$ values in ratio-only runs thus reflect overparameterization rather than evidence for DDE.

These results underscore that the apparent variations of $(\oo,\,\wa)$ across different BAO bases and prior widths primarily trace the same degeneracy ridge rather than indicating genuine dark-energy evolution.  The location of the posterior along this ridge is controlled mainly by the adopted prior and basis geometry, while the underlying physical inference remains invariant.  Absolute-scale anchoring through observables such as $D_V/r_d$ is therefore essential for reliable BAO-based cosmological inference, and redundant combinations of correlated data should be avoided to maintain statistical consistency.  When BAO data are analyzed in isolation, they provide purely geometric constraints and yield a robust, prior-independent result of $\wpp\simeq-0.9\pm0.1$, fully consistent with a cosmological constant.  However, combining BAO with external probes such as the CMB or supernovae imposes strong priors on $\Omegam$, $h$, and $r_d$, which can compress the posterior along the $(\oo,\,\wa)$ degeneracy ridge.  This compression often drives $\wa$ toward negative values, creating the illusion of a time-varying dark-energy equation of state.  Such apparent evolution does not reflect new information in the data but instead arises from the statistical geometry of the joint likelihood.  The BAO-only constraint on $\wpp$ thus provides a physically meaningful baseline against which the influence of external priors can be quantitatively assessed.

In summary, the analysis presented here demonstrates that the DESI~DR2 BAO likelihood provides no statistically significant evidence for dynamical dark energy.  Apparent deviations from $\omega=-1$ arise from degeneracy geometry and prior sensitivity rather than from genuine evolution in $\omega(a)$.  The results reinforce the internal consistency of $\LCDM$ under BAO-only tests and establish a transparent, self-consistent framework for evaluating dark-energy models with future DESI data releases and next-generation surveys.


\end{document}